\documentclass[letterpaper]{article}

\usepackage[T1]{fontenc}

\usepackage{geometry}
\usepackage{setspace}
\usepackage{multirow}
\usepackage[doi = true, articletitle = true]{achemso}

\usepackage{graphicx, subcaption}
\usepackage{float}
\newfloat{scheme}{htbp}{los}
\floatname{scheme}{Scheme}
\floatname{chart}{Chart}
\newfloat{graph}{htbp}{loh}

\usepackage{chemformula} 
\usepackage[version = 4]{mhchem} 

\usepackage{booktabs}    
\usepackage{amsmath, amssymb, bm}     
\usepackage{siunitx}     
\usepackage[hidelinks]{hyperref}
\usepackage{parskip}

\usepackage{authblk}
\author[1]{Sayan Bhowmik}
\author[1]{Andrew J. Medford}
\author[1, 2]{Phanish Suryanarayana*}
\affil[1]{College of Engineering, Georgia Institute of Technology, Atlanta, GA 30332, USA}
\affil[2]{College of Computing, Georgia Institute of Technology, Atlanta, GA 30332, USA}

\title{Bulk Boundary Condition for Surface Calculations in Density Functional Theory}

\date{*Email: phanish.suryanarayana@ce.gatech.edu}

\begin{document}

\maketitle

\begin{abstract}
We present a bulk boundary condition formalism for surface calculations in Kohn--Sham density functional theory. The approach exploits the nearsightedness of electronic interactions in real space to restrict the calculation to a localized surface region. Within this region, the electron density is evaluated by leveraging the decay of the density matrix, with bulk values imposed on the density and electrostatic potential in the interior, and the electrostatic potential solved subject to bulk boundary conditions. The energy and atomic forces are computed using density-matrix-based expressions. Through representative calculations of surface and adsorption energies, we demonstrate the accuracy and efficiency of the proposed formalism.
\end{abstract}



\begin{figure}[h]
    \centering
    \includegraphics[width=0.6\linewidth]{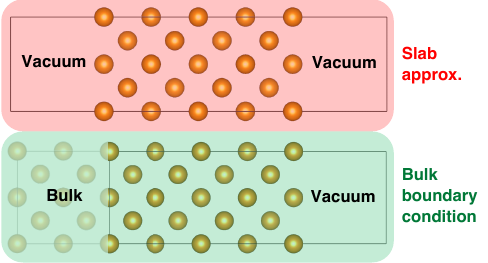}
    \caption{TOC Graphic}
    \label{fig:TOC}
\end{figure}

%
%

Surfaces govern the functional behavior of materials in catalysis~\cite{HAMMER200071, Norskov2009, norskov2014fundamental}, electrochemistry~\cite{Greeley2006, Rossmeisl2006, Skulason2007}, corrosion~\cite{GREELEY20075829, todorova2014}, and electronic devices~\cite{vanDeWalle1987, Dardzinski_2022}, where properties such as adsorption energetics~\cite{HAMMER200071, chen2021}, work functions~\cite{singhMiller2009, Fall_1999}, and charge transfer~\cite{neugebauer1992, Liu2023} are controlled by the local electronic structure within the surface region. As extended defects that break translational symmetry, surfaces give rise to under-coordinated atoms~\cite{somorjai1994, zangwill1988}, charge redistribution~\cite{neugebauer1992, BeFriedel1997}, band bending~\cite{SiSlowConvSlab2017, vanDeWalle1987}, and structural relaxation or reconstruction~\cite{singhMiller2009, Schultz2021}. These structural and electronic modifications carry direct scientific and technological consequences, as it is the surface that mediates interactions with adsorbates or adjacent phases, thereby controlling catalytic activity~\cite{HAMMER200071, Norskov2009, chen2021}, corrosion resistance\cite{GREELEY20075829, todorova2014}, and device performance~\cite{vanDeWalle1987, Dardzinski_2022}. Because these effects originate in subtle rearrangements of the electron density and are often not captured by empirical or bulk-derived models\cite{justo1998,mattsson2002}, a quantitative and predictive description of surface behavior requires electronic structure calculations that resolve the surface explicitly.

Kohn--Sham density functional theory (DFT)~\cite{hohenberg1964inhomogeneous, kohnSham1965} has become a cornerstone of materials and chemical sciences research, owing to its predictive power and favorable computational cost-accuracy tradeoff relative to other such ab initio approaches~\cite{burkePerspective2012, beckeFifty2014}. However, Kohn--Sham calculations incur substantial computational costs, scaling cubically with system size, which limits the range of systems that can be studied. Surface calculations present an additional and more fundamental challenge: the semi-infinite physical system must be represented within a finite computational domain. Since the Kohn--Sham formalism is expressed in terms of orbitals, which are global quantities defined over the entire system, restricting the calculation to a finite domain near the surface requires boundary conditions on these orbitals at the bulk-facing edge. However, the associated eigenproblem does not admit such a truncation: in general, only Dirichlet or Bloch-periodic boundary conditions can be prescribed, neither of which is appropriate at an arbitrary artificial boundary within an otherwise periodic bulk.  Even if suitable conditions could be identified, the number of orbitals grows with system size, and so the semi-infinite system entails solving for an infinite number of orbitals, even if only the near-surface region is of interest. A further complication arises from the electrostatics, which can be long-ranged; imposing physically appropriate boundary conditions on the electrostatic potential of a semi-infinite system likewise requires an appropriate real-space treatment. Neither requirement can be met in the widely used planewave DFT calculations \cite{kresse1996,ABINIT,CASTEP,Espresso}, where the Fourier basis imposes artificial periodicity on the system. Real-space DFT methods~\cite{chelikowsky1994, ghosh2017sparcI, ghosh2017sparc2, MOTAMARRI2020106853} partially address these limitations, as they admit Dirichlet boundary conditions; they are moreover well suited to high-performance computing~\cite{xu2021sparc,zhang2024sparc,gavini2022roadmap} and the development of linear-scaling methods~\cite{shimojo2001linear,SURYANARAYANA2013182}, enabling access to significantly larger system sizes. However, the fundamental challenge of representing a semi-infinite surface within a finite computational domain remains unresolved.

The standard approach for surface calculations is the slab formalism, in which the surface is represented by a finite number of atomic layers sandwiched between vacuum regions~\cite{gross2009theoretical, norskov2014fundamental, sholl2022density}. While nearly universally used, this formalism introduces two surfaces by construction, requiring management of the interactions between them as well as dipole corrections to remove the spurious electric fields arising for asymmetric slabs~\cite{neugebauer1992, makov1995}. Moreover, convergence of surface properties with slab thickness can be slow~\cite{Fiorentini_1996, Fall_1999, singhMiller2009, Martsinovich2010}, both for metals~\cite{metalSlowConv2013} and for band-gap containing systems~\cite{GeSiSlowConvSlab2012, SiSlowConvSlab2017}. Frozen-density embedding methods provide a route to such calculations that avoids finite slabs, treating the surface region self-consistently within the frozen density of its semi-infinite environment~\cite{cohen2003, Inglesfield1988, layerKKR1989, surfaceKKR1994, Inglesfield_1981}. However, these schemes couple the subsystem to its environment through an approximate (typically orbital-free) non-additive kinetic energy functional, which limits their accuracy. The Green's function method instead incorporates the semi-infinite bulk exactly, coupling the surface region to it via the Dyson equation and computing the density matrix through a contour integral~\cite{sancho_1984, sancho_1985, Inglesfield1988, layerKKR1989, sgf1991, surfaceKKR1994, sgfLMTO1992}. However, this approach is computationally expensive, typically employs localized basis sets that are not systematically improvable, and requires specialized code infrastructure not available in standard DFT codes. Consequently, a formulation of Kohn--Sham DFT for surfaces that efficiently and faithfully treats the semi-infinite system while retaining the systematic convergence and computational structure of standard real-space DFT is highly desirable, and motivates the present work.

In this letter, we develop a bulk boundary condition formalism that brings the semi-infinite nature of surfaces into Kohn--Sham DFT, confining calculations to a localized region containing the surface. We demonstrate its accuracy and efficiency through surface and adsorption energy calculations for representative systems.


In spin-unpolarized Kohn--Sham DFT with local/semilocal exchange-correlation, real-space electrostatics, and pseudopotentials, the nonlinear eigenproblems governing the electronic ground state for extended systems take the form \cite{ghosh2017sparc2}:
\begin{align}
\left(\mathcal{H}_{\bm{k}} \equiv -\frac{1}{2}\nabla^2 + V_{\mathrm{xc}} + \phi + \mathcal{V}_{nl,\bm{k}} \right) \psi_{n \bm{k}}
= \lambda_{n \bm{k}} \psi_{n \bm{k}} \, ,
\end{align}
where $\bm{k}$ are the Brillouin-zone wavevectors, $\mathcal{H}_{\bm{k}}$ are the Hamiltonians, $\psi_{n\bm{k}}$ are the orbitals, $\lambda_{n\bm{k}}$ are the corresponding eigenvalues, $V_{\mathrm{xc}}$ is the exchange-correlation potential, $\phi$ is the electrostatic potential, and $\mathcal{V}_{nl,\bm{k}}$ is the nonlocal pseudopotential operator. The electron density, which couples the different eigenproblems, is given by:
\begin{align}
\rho = 2 \sum_{\bm{k}} w_{\bm{k}} \sum_n f_{n\bm{k}} |\psi_{n\bm{k}}|^2 \,,
\end{align}
where $w_{\bm{k}}$ the Brillouin zone weights and $f_{n\bm{k}}$ the orbital occupations, calculated using the Fermi-Dirac distribution, while enforcing the constraint on the number of electrons through the Fermi level $\mu$. The exchange-correlation potential $V_{\mathrm{xc}}$ depends on the density alone in local density approximation, and on the density and its gradient in generalized gradient approximation. The electrostatic potential is given by the solution of the Poisson equation \cite{suryanarayana2014augmented, ghosh2016higher}: 
\begin{align}
-\frac{1}{4\pi}\nabla^2\phi = \rho + b \, ,
\end{align}
where $b$ is the total ionic pseudocharge density. The nonlocal pseudopotential operator $\mathcal{V}_{nl,\bm{k}}$ is an integral operator that is determined by the atomic positions along with appropriate Bloch-periodic mapping. For a semi-infinite surface, the orbitals $\psi_{n\bm{k}}$ satisfy zero-Dirichlet boundary conditions at the vacuum boundary and Bloch-periodic boundary conditions in the lateral directions, while the electrostatic potential $\phi$ satisfies a Dirichlet condition at the vacuum boundary and periodic boundary conditions in the lateral directions. Although the real-space formulation of electrostatics naturally accommodates a bulk boundary condition on $\phi$ in the interior, the same does not hold for $\psi_{n\bm{k}}$: as discussed above, the corresponding boundary conditions are not known a priori, and a direct solution of the semi-infinite problem would in any case be computationally intractable.

The Kohn--Sham problem above can be equivalently reformulated in terms of the density matrix:
\begin{align}
\mathcal{D}(\bm{x},\bm{x'}) = \sum_{\bm{k}} w_{\bm{k}} \sum_n f_{n \bm{k}} \psi_{n \bm{k}}(\bm{x}) \psi^*_{n \bm{k}}(\bm{x'}) \,,
\end{align}
where $(\cdot)^{*}$ denotes the complex conjugate. The electron density can then be written as:
\begin{align}
\rho(\bm{x}) = 2 \mathcal{D}(\bm{x},\bm{x}) \,,
\end{align}
with the Fermi level $\mu$ calculated from the constraint that the integrated electron density equals the number of electrons. Once the electronic ground state has been determined, quantities such as the energy, atomic forces, and stresses can be evaluated using the density matrix \cite{SURYANARAYANA2018288, PRATAPA201696, SURYANARAYANA2013182, SURYANARAYANA201338}. Unlike the orbitals and eigenvalues, which are global quantities, the real-space density matrix is a local quantity that decays exponentially for insulators and metallic systems at finite smearing \cite{goedeckerDecay1998, AriasDecay1999, razoukDecay2013, SURYANARAYANA2017146}.Furthermore, only the values of the density matrix with one argument restricted to the domain of interest are required to evaluate the quantities therein \cite{SURYANARAYANA2018288, PRATAPA201696, SURYANARAYANA201338}. These features enable calculations to be restricted to the near-surface region and make the density-matrix formulation of real-space DFT suitable for developing the bulk boundary condition formalism, as discussed next for surfaces.

\begin{figure}[h!]
    \centering
    \includegraphics[width=0.75\textwidth]{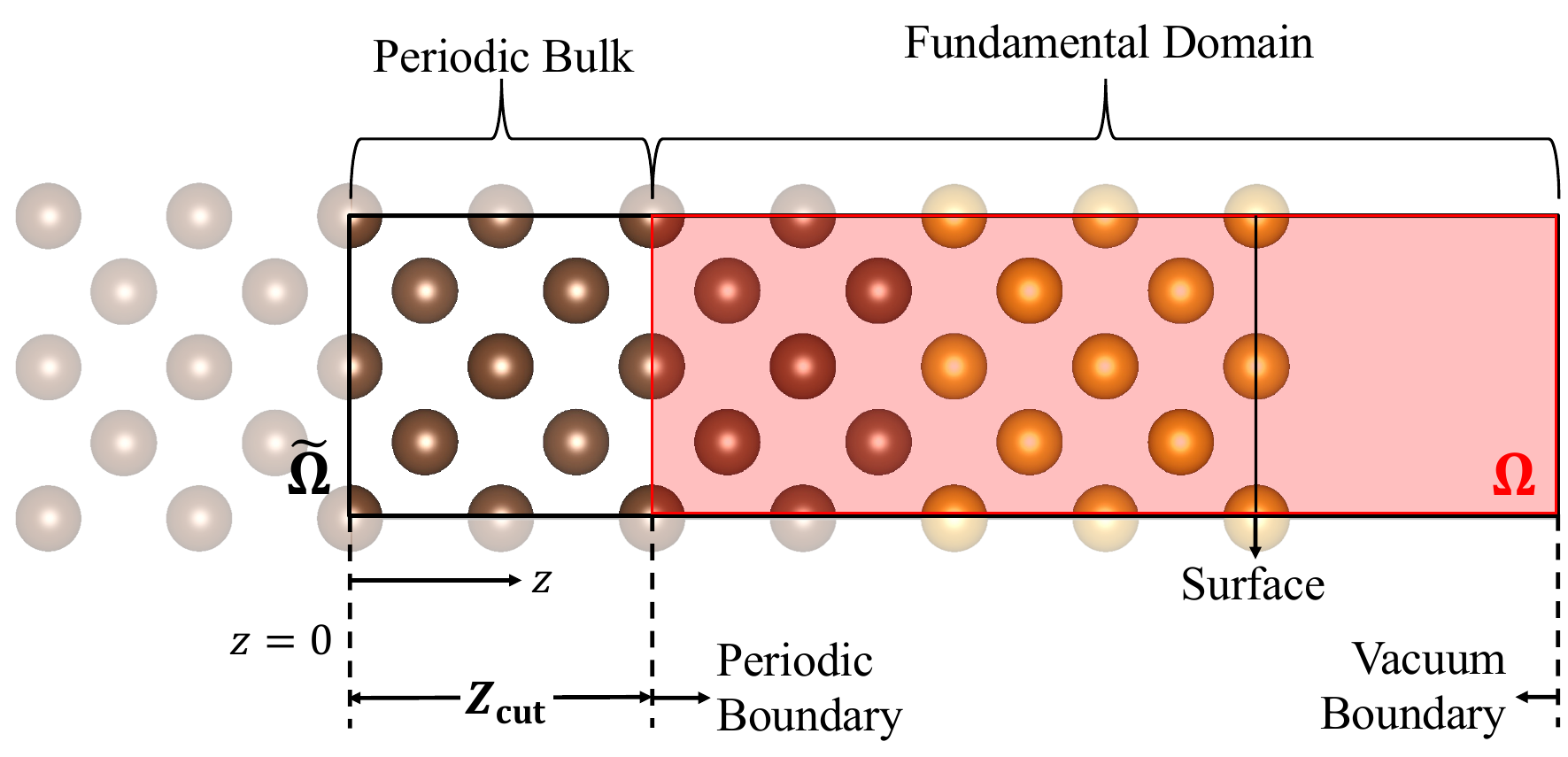}
    \caption{Schematic of the bulk boundary condition formalism for surface calculations. The fundamental domain $\Omega$ contains the surface with vacuum to the right, and is embedded within the extended domain $\widetilde{\Omega}$, which includes a bulk region of thickness $Z_{\mathrm{cut}}$ to the left. The electron density and energy are computed on $\Omega$, atomic forces are computed for atoms within $\Omega$, while the electrostatic potential is solved on the larger domain $\widetilde{\Omega}$. The electron density and electrostatic potential are constrained to their bulk values in the regions $z < Z_{\mathrm{cut}}$ and $z < 0$, respectively. Darker-colored atoms are held fixed at their bulk positions, while lighter-colored atoms are free to relax.}
    \label{fig:Domain}
\end{figure}

Consider the setup illustrated in Fig.~\ref{fig:Domain}, where the fundamental domain $\Omega$ contains the surface region of interest with vacuum to its right, and $\widetilde{\Omega}$ denotes the extended domain obtained by augmenting $\Omega$ with a bulk region of thickness $Z_{\mathrm{cut}}$ in the surface-normal direction $z$; for simplicity, both $\Omega$ and $\widetilde{\Omega}$ are taken to be an integer number of unit cells in thickness. Here, $z = 0$ marks the boundary between $\widetilde{\Omega}$ and the bulk, and $z = Z_{\mathrm{cut}}$ the periodic boundary. The electron density is constrained to its bulk value for $z < Z_{\mathrm{cut}}$, and the electrostatic potential for $z < 0$; the latter is solved on the larger domain $\widetilde{\Omega}$, since the surface-induced perturbation decays more slowly in the potential than in the density. The electron density and energy are computed on $\Omega$, and atomic forces for atoms within it. Atoms in the periodic bulk region and in the first unit cell of $\Omega$ to the right of the periodic boundary at $z = Z_{\mathrm{cut}}$ are held fixed at their bulk positions, while atoms closer to the surface within $\Omega$ are free to relax. In principle, all atoms within $\Omega$ may be relaxed; however, the current strategy reduces mismatch and prevents electrons from entering or leaving the computational domain.

Consider the following density matrix defined on $\widetilde{\Omega}$:
\begin{align}
\widetilde{\mathcal{D}}(\bm{x},\bm{x'}) = \sum_{\bm{k}} w_{\bm{k}} \sum_n \tilde{f}_{n \bm{k}} \widetilde{\psi}_{n \bm{k}}(\bm{x}) \widetilde{\psi}^*_{n \bm{k}}(\bm{x'}) \,,
\end{align}
where $\widetilde{\psi}_{n \bm{k}}$ are the orbitals of the truncated Kohn--Sham Hamiltonian $\widetilde{\mathcal{H}}_{\bm{k}}$ on $\widetilde{\Omega}$:
\begin{align}
\left(\widetilde{\mathcal{H}}_{\bm{k}} \equiv -\frac{1}{2}\nabla^2 + \widetilde{V}_{\mathrm{xc}} + \widetilde{\phi} + \widetilde{\mathcal{V}}_{nl,\bm{k}} \right) \widetilde{\psi}_{n \bm{k}} = \widetilde{\lambda}_{n \bm{k}} \widetilde{\psi}_{n \bm{k}} \,,
\end{align}
with $\widetilde{\lambda}_{n \bm{k}}$ and $\tilde{f}_{n \bm{k}}$ the corresponding eigenvalues and occupations, respectively. The eigenfunctions $\widetilde{\psi}_{n \bm{k}}$ are subject to zero Dirichlet boundary conditions at the vacuum boundary and at $z=0$, and Bloch boundary conditions in the lateral directions. The electron density $\widetilde{\rho}$ on $\widetilde{\Omega}$ is defined as:
\begin{align}
\widetilde{\rho}(\bm{x}) =
\begin{cases}
2\widetilde{\mathcal{D}}(\bm{x},\bm{x}) & z > Z_{\mathrm{cut}} \\
\rho_{\mathrm{bulk}}(\bm{x}) & z \leq Z_{\mathrm{cut}} 
\end{cases} \,,
\end{align}
where $\rho_{\mathrm{bulk}}$ is the density corresponding to a 3D bulk calculation, and the Fermi level $\mu$ is determined by the integral constraint on the number of electrons $N_{\rm electrons}$ in $\Omega$: $\int_{\Omega} \widetilde{\rho}(\bm{x}) \, d \bm{x} = N_{\rm electrons}$, while employing Fermi-Dirac smearing for the occupations $\tilde{f}_{n \bm{k}}$. The exchange-correlation potential $\widetilde{V}_{\mathrm{xc}}$ represents the restriction of the exchange-correlation potential for the semi-infinite surface to $\widetilde{\Omega}$. The electrostatic potential $\widetilde{\phi}$ is the solution of the Poisson equation on $\widetilde{\Omega}$:
\begin{align}
-\frac{1}{4\pi}\nabla^2 \widetilde{\phi} = \widetilde{\rho} + \widetilde{b} \,,
\end{align}
where $\widetilde{b}$ is the restriction of the total ionic pseudocharge density for the semi-infinite surface to $\widetilde{\Omega}$. The electrostatic potential $\widetilde{\phi}$ is subject to zero Dirichlet boundary conditions at the vacuum boundary, periodic boundary conditions in the lateral directions, and a bulk boundary condition at $z=0$: $\widetilde{\phi} = \phi_{\mathrm{bulk}}$, where $\phi_{\mathrm{bulk}}$ is the electrostatic potential corresponding to a 3D bulk calculation. Since $\phi_{\mathrm{bulk}}$ is determined only up to an additive constant, this constant is fixed by enforcing continuity between $\widetilde{\phi}$ and $\phi_{\mathrm{bulk}}$ at $z=0$. The nonlocal pseudopotential operator $\widetilde{\mathcal{V}}_{nl,\bm{k}}$ represents the restriction of the nonlocal pseudopotential operator for the semi-infinite surface to $\widetilde{\Omega}$, corresponding to zero Dirichlet boundary conditions on $\widetilde{\psi}_{n \bm{k}}$ for $z\leq 0$.

Once the electronic ground state has been determined, the electronic free energy of the system in $\Omega$:
\begin{align}
    E(\bm{R}) =& -\sum_{\bm{k}} w_{\bm{k}} \sum_n \tilde{f}_{n \bm{k}} \int_{\Omega} \widetilde{\psi}^*_{n \bm{k}}(\bm{x}) \nabla^2 \widetilde{\psi}_{n \bm{k}}(\bm{x}) \, d\bm{x} + \int_{\Omega} \varepsilon_{\mathrm{xc}} \big( \widetilde{\rho}(\bm{x}),\nabla \widetilde{\rho}(\bm{x}) \big)\widetilde{\rho}(\bm{x}) \, d\bm{x} \nonumber \\
    &+ \frac{1}{2} \int_{\Omega}  \big( \widetilde{\rho}(\bm{x}) + \widetilde{b}(\bm{x}, \bm{R}) \big) \widetilde{\phi}(\bm{x}, \bm{R}) \, d\bm{x}  + E_{\mathrm{sc}}(\bm{R}) \nonumber \\
    &+ 2 \sum_{\bm{k}} w_{\bm{k}} \sum_n \tilde{f}_{n \bm{k}} \int_{\Omega} \widetilde{\psi}^*_{n \bm{k}}(\bm{x}) \int_{\widetilde{\Omega}} \widetilde{\mathcal{V}}_{nl, \bm{k}}(\bm{x},\bm{x'}) \widetilde{\psi}_{n \bm{k}} (\bm{x'})\, d\bm{x'} \, d\bm{x} \nonumber \\
    & + 2 \sigma \sum_{\bm{k}} w_{\bm{k}} \sum_n \big( \tilde{f}_{n \bm{k}} \log (\tilde{f}_{n \bm{k}}) + (1 - \tilde{f}_{n \bm{k}} ) \log (1 - \tilde{f}_{n \bm{k}}) \big) \int_{\Omega} \big| \widetilde{\psi}_{n \bm{k}} (\bm{x}) \big|^2 \, d\bm{x} \,,
\end{align}
where $\bm{R} = \{\bm{R}_1, \bm{R}_2, \ldots, \bm{R}_{n_{\rm atoms}} \}$ represents the atomic configuration, $\varepsilon_{\mathrm{xc}}$ is the exchange-correlation energy per electron, $E_{sc}$ is the self and overlap  energy correction associated with the ionic pseudocharge densities \cite{ghosh2016higher, ghosh2017sparc2}, and $\sigma$ is the smearing parameter.  The terms in the energy correspond, in order, to the electronic kinetic energy, the exchange-correlation energy, the electrostatic energy (third and fourth terms), the nonlocal pseudopotential energy, and the electronic entropy energy.

The Hellmann-Feynman forces for atoms in $\Omega$:
\begin{align}
    \bm{f}_I &= \sum_{I'} \int_{\widetilde{\Omega}} b_{I'}\big(\bm{x},\bm{R}_{I'}\big) \nabla \widetilde{\phi}(\bm{x},\bm{R}) \, d\bm{x} + \bm{f}^{\mathrm{sc}}_{I} \nonumber \\
    &- 2\sum_{\bm{k}} w_{\bm{k}} \sum_{I'}\sum_n \tilde{f}_{n\bm{k}} \int_{\widetilde{\Omega}} \int_{\widetilde{\Omega}} \widetilde{\psi}^*_{n\bm{k}}(\bm{x})\, \widetilde{\mathcal{V}}_{nl,\bm{k}}^{I'}(\bm{x},\bm{x'})\, \nabla \widetilde{\psi}_{n\bm{k}}(\bm{x'})\, d\bm{x'}\, d\bm{x} \,,
\end{align}
where the summation index $I'$ runs over atom $I$ and its periodic images in the lateral directions, $b_{I'}$ represents the ionic pseudocharge of the associated atom, and $\bm{f}^{\mathrm{sc}}_{I}$ represents the force associated with the self and overlap energy correction. The first term in the force is the electrostatic contribution, and the second the nonlocal pseudopotential contribution. This expression, derived following previous work \cite{PRATAPA201696, SURYANARAYANA2018288}, involves integrals over $\widetilde{\Omega}$, the ionic pseudocharges and nonlocal projectors extending into this region.

The present formalism exploits the decay of the density matrix to compute the electron density, energy, and atomic forces in $\Omega$. The density matrix $\widetilde{\mathcal{D}}(\bm{x}, \bm{x'})$ is a controlled approximation to the exact one for $\bm{x} \in \Omega$, converging to it as $Z_{\mathrm{cut}}$ is increased. Though the discussion has focused on a clean surface, the formalism applies equally in the presence of adsorbates. While restricted here to non-spin-polarized calculations, extension to spin polarization is straightforward, requiring separate density matrices for the two spin channels. Extension to nonlocal exchange-correlation functionals such as hybrids, though expressible in terms of the density matrix, will require methodology to incorporate nonlocal exact exchange; extension to surfaces with spontaneous bulk polarization, and to charged adsorbates on them, will require appropriate generalization of the electrostatics.

The present formalism is related to the infinite-cell variant of the spectral quadrature method~\cite{SURYANARAYANA2018288, PRATAPA201696}, in which bulk boundary conditions enable DFT calculations without Brillouin zone sampling or large supercells. That method, however, exploits translational symmetry and is therefore inapplicable to defects, including surfaces; the present formalism extends this concept of environment modeling in Kohn--Sham DFT to semi-infinite surface calculations. The approach also shares similarities with the frozen-density embedding methods discussed above~\cite{cohen2003, Inglesfield1988, layerKKR1989, surfaceKKR1994, Inglesfield_1981}. Whereas those schemes partition the system into subsystems coupled through an approximate non-additive kinetic energy functional, the present approach involves no such partitioning: the surface region is described by the same Kohn--Sham problem as the full semi-infinite system, with the bulk entering exactly through the boundary conditions. Finally, while the Green's function method~\cite{sancho_1984, sancho_1985, Inglesfield1988, layerKKR1989, sgf1991, surfaceKKR1994, sgfLMTO1992} also incorporates the semi-infinite bulk exactly, the present formalism is significantly more efficient, compatible with general real-space bases, and implementable within existing Kohn--Sham infrastructure.


We implemented the bulk boundary condition formalism in M-SPARC~\cite{xu2020m, zhang2023version}, the MATLAB counterpart of the SPARC~\cite{xu2021sparc, zhang2024sparc} electronic structure code. In particular, we employ a uniform real-space grid with high-order centered finite differences to approximate derivatives, and the trapezoidal rule for numerical integration. The grid spacing is chosen to resolve the bulk unit cell; since $\Omega$ and $\widetilde{\Omega}$ each contain an integer number of unit cells, a plane of nodes automatically coincides with $z = Z_{\mathrm{cut}}$, ensuring a consistent partitioning of the discretization between $\Omega$ and the bulk region. The bulk values of the electron density and electrostatic potential are obtained from a separate calculation on the 3D crystal, performed on a grid commensurate with that of the surface calculation. Zero-Dirichlet boundary conditions are enforced by setting out-of-domain values to zero~\cite{ghosh2017sparcI}, while Bloch-periodic boundary conditions are enforced by mapping out-of-domain indices back into the grid with the appropriate Bloch phase factor~\cite{ghosh2017sparc2}. The nonzero Dirichlet boundary conditions are enforced by evaluating stencil contributions that extend beyond the domain boundary using the prescribed boundary values, and transferring them to the right-hand side of the linear system~\cite{ghosh2017sparcI}. The electronic ground state is computed using the self-consistent field (SCF) method.

We first demonstrate the accuracy and efficiency of the proposed formalism through calculations of the (100) surface energies of face-centered cubic aluminum (Al), diamond cubic silicon (Si), and diamond cubic carbon (C)---prototypical metallic, semiconducting, and insulating systems, respectively. In all calculations, the computational domain spans a single unit cell in the lateral directions, with a vacuum region of one unit cell. We employ the Perdew--Burke--Ernzerhof (PBE)~\cite{perdew1996generalized} generalized gradient approximation for the exchange-correlation functional and optimized norm-conserving Vanderbilt (ONCV)~\cite{hamann2013optimized} pseudopotentials from the SPMS set~\cite{shojaei2023soft}. The unit cells correspond to equilibrium lattice constants of $a_{\rm lat}=$ 7.63, 10.33, and 6.75~bohr for Al, Si, and C, respectively. For Al, Fermi--Dirac smearing of 0.1~eV is used. Twelfth-order finite differences are used throughout, with grid spacings of 0.254, 0.295, and 0.241~bohr for Al, Si, and C, respectively. The surface calculations employ $10 \times 10$, $4 \times 4$, and $4 \times 4$ k-point grids, while the corresponding bulk calculations for the 3D bulk electron density and electrostatic potential employ $10 \times 10 \times 10$, $4 \times 4 \times 8$, and $4 \times 4 \times 8$ grids. For geometry optimization, we employ a tolerance of $10^{-4}$~ha/bohr on the atomic forces. These and other numerical parameters are chosen such that the surface energies are converged to within 0.025~eV/surface atom. We compare the results with those from the slab formalism, computed using SPARC~\cite{xu2021sparc, zhang2024sparc} with identical simulation parameters.

\begin{figure}[h!]
    \centering
    \begin{subfigure}[b]{0.38\linewidth}
        \centering
        \includegraphics[width=\textwidth]{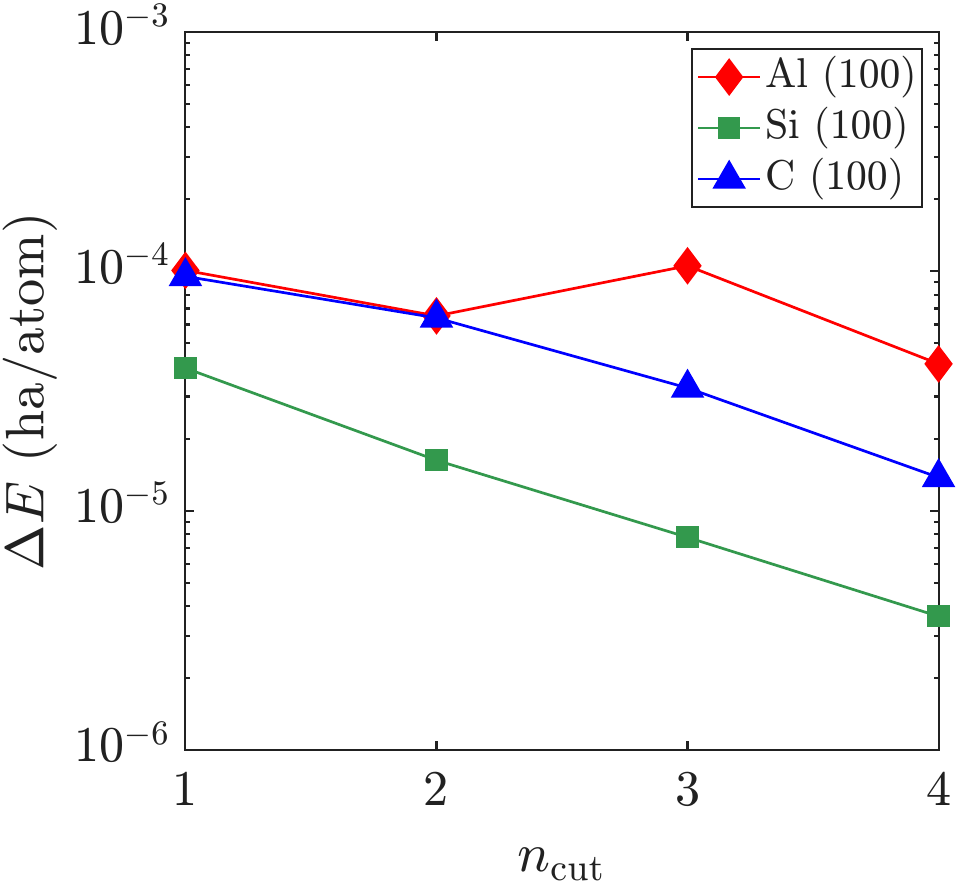}
        \caption{Energy}
        \label{fig:E_convergence}
    \end{subfigure}
    \hspace{1em}
    \begin{subfigure}[b]{0.38\linewidth}
        \centering
        \includegraphics[width=\textwidth]{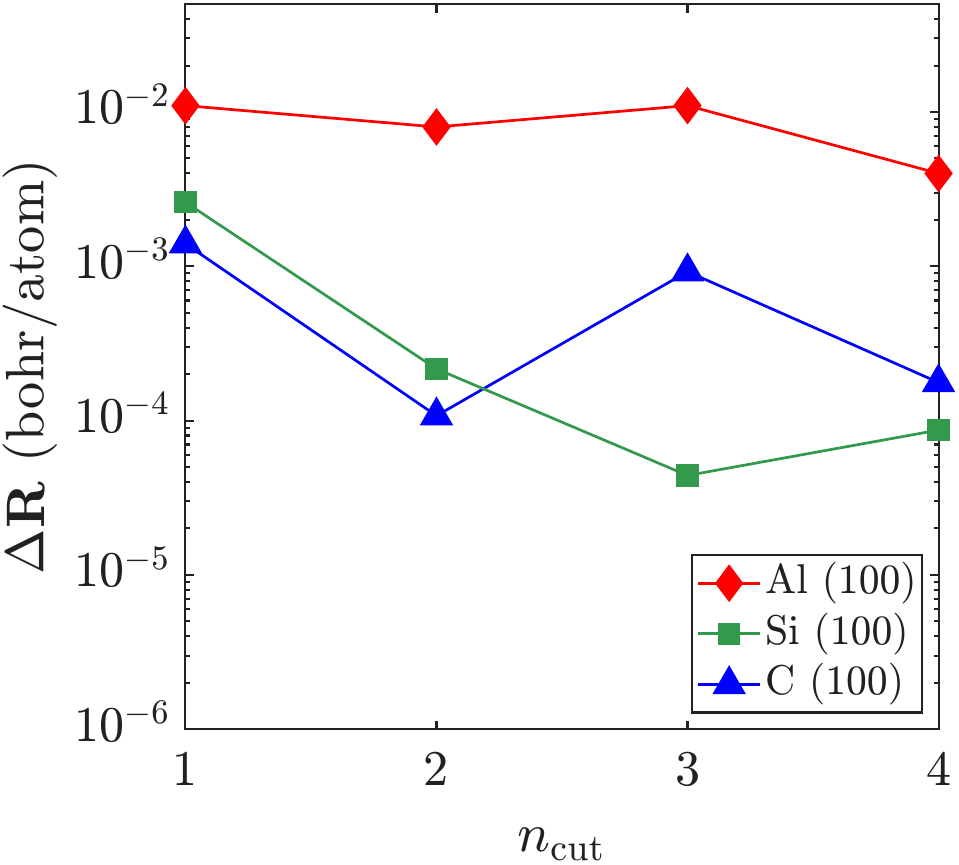}
        \caption{Atomic positions}
        \label{fig:Position_convergence}
    \end{subfigure}
    \caption{Convergence of (a) the total energy and (b) the relaxed atomic positions with $Z_{\mathrm{cut}} = n_{\mathrm{cut}} \, a_{\mathrm{lat}}$, relative to the results at $Z_{\mathrm{cut}} = 5 \, a_{\mathrm{lat}}$. The domain $\Omega$ consists of two unit cells, and the  position error is measured as the norm of the position difference per relaxed atom.}
    \label{fig:E_and_pos_conv}
\end{figure}

Relative to standard DFT calculations, the proposed formalism introduces one additional parameter, $Z_{\mathrm{cut}}$, with respect to which convergence must be verified. We therefore first examine the convergence of the total energy and relaxed atomic positions with $Z_{\mathrm{cut}}$. The results are presented in Fig.~\ref{fig:E_and_pos_conv} for a system with two unit cells in $\Omega$, using the values at $Z_{\mathrm{cut}} = 5\,a_{\mathrm{lat}}$ as reference.  The convergence with $Z_{\mathrm{cut}}$ is generally non-monotonic, consistent with the literature for 3D periodic calculations~\cite{SURYANARAYANA2013182,PRATAPA201696}. For Al, the non-monotonicity is more pronounced and longer ranged due to the Friedel oscillations characteristic of metallic systems~\cite{BeFriedel1997, Schultz2021}. The rate of convergence reflects the decay of the density matrix, which is faster for larger band gaps in insulating systems and, analogously, for larger smearing in metallic systems~\cite{SURYANARAYANA2013182}. Accordingly, convergence is fastest for C, which has the largest band gap, followed by Si, with Al the slowest. The norm of the difference in atomic positions, computed over the relaxed atoms and normalized by their number, converges to within $10^{-3}$ bohr/atom for Si and C, and $10^{-2}$ bohr/atom for Al, as shown in Fig.~\ref{fig:Position_convergence}. These results confirm that $Z_{\mathrm{cut}} = a_{\mathrm{lat}}$ suffices for practical accuracy in Si and C, while Al requires slightly greater depth. In all subsequent calculations, we nevertheless employ conservative values of $Z_{\mathrm{cut}} = 4\,a_{\mathrm{lat}}$ for Al and C, and $3\,a_{\mathrm{lat}}$ for Si, so as to verify the accuracy of the formalism.

\begin{figure}[h!]
    \centering
    \begin{subfigure}[b]{0.48\linewidth}
        \centering
        \includegraphics[width=0.75\textwidth]{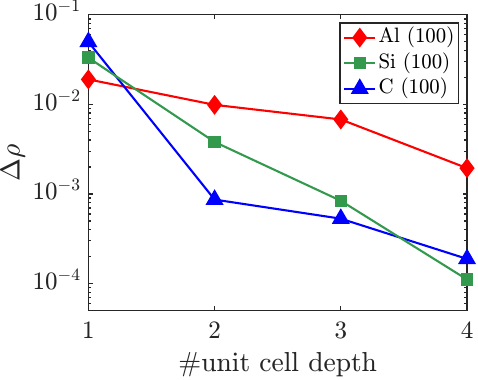}
        \caption{Convergence of the electron density}
        \label{fig:Rho_convergence}
    \end{subfigure}
    \hfill
    \begin{subfigure}[b]{0.48\linewidth}
        \centering
        \includegraphics[width=0.75\textwidth]{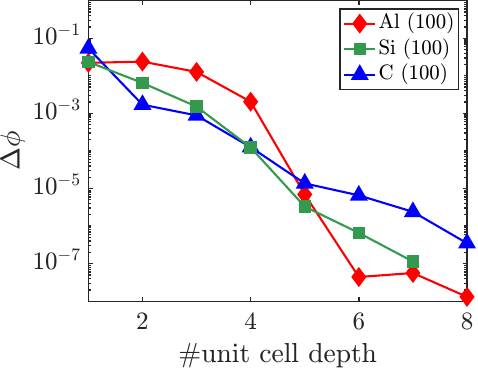}
        \caption{Convergence of the electrostatic potential}
        \label{fig:Phi_convergence}
    \end{subfigure}

    \vspace{1em} 
    
    \begin{subfigure}[b]{0.48\linewidth}
        \centering
        \includegraphics[width=\textwidth]{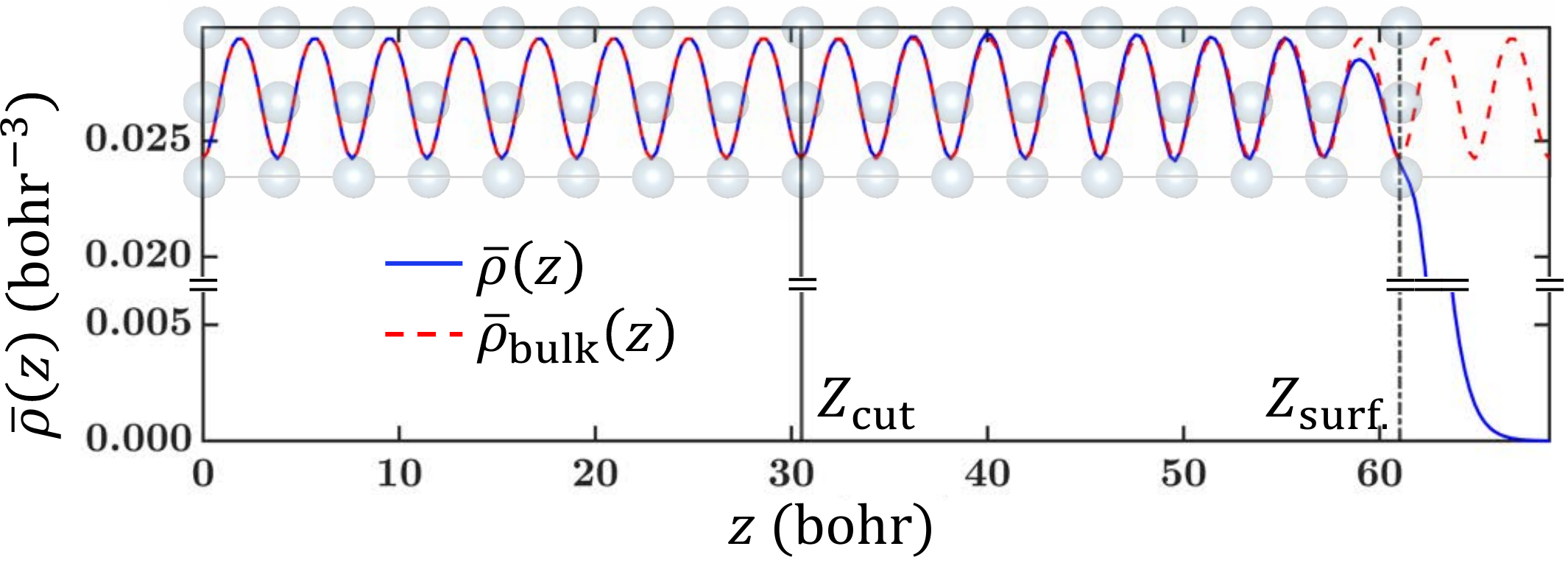}
        \caption{$\bar{\rho}(z)$ for Al}
        \label{fig:Rho_Al}
    \end{subfigure}
    \hfill
    \begin{subfigure}[b]{0.48\linewidth}
        \centering
        \includegraphics[width=\textwidth]{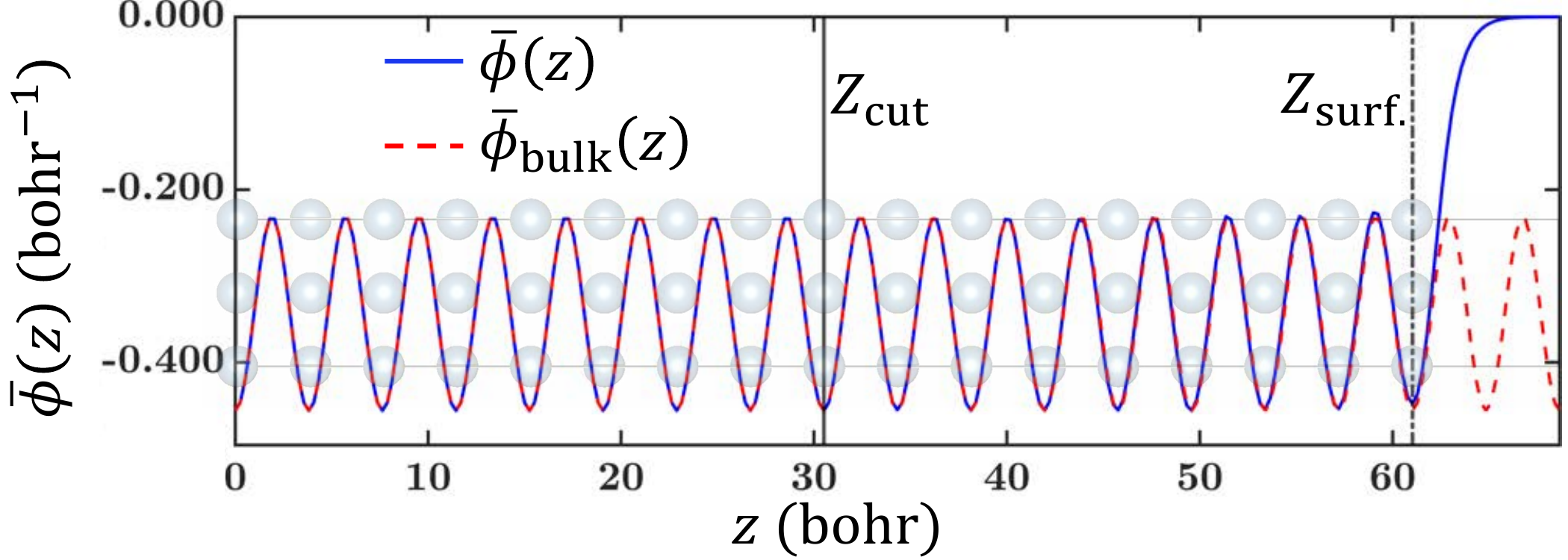}
        \caption{$\bar{\phi}(z)$ for Al}
        \label{fig:Phi_Al}
    \end{subfigure}

    \vspace{1em} 

    \begin{subfigure}[b]{0.48\linewidth}
        \centering
        \includegraphics[width=\textwidth]{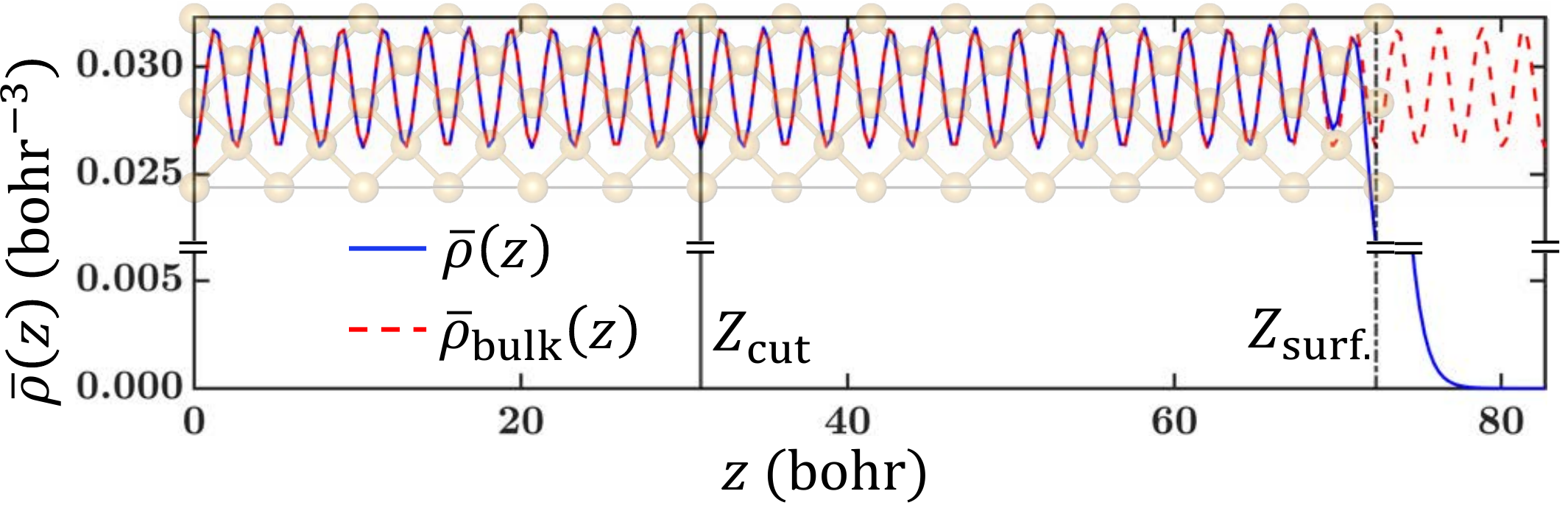}
        \caption{$\bar{\rho}(z)$ for Si}
        \label{fig:Rho_Si}
    \end{subfigure}
    \hfill
    \begin{subfigure}[b]{0.48\linewidth}
        \centering
        \includegraphics[width=\textwidth]{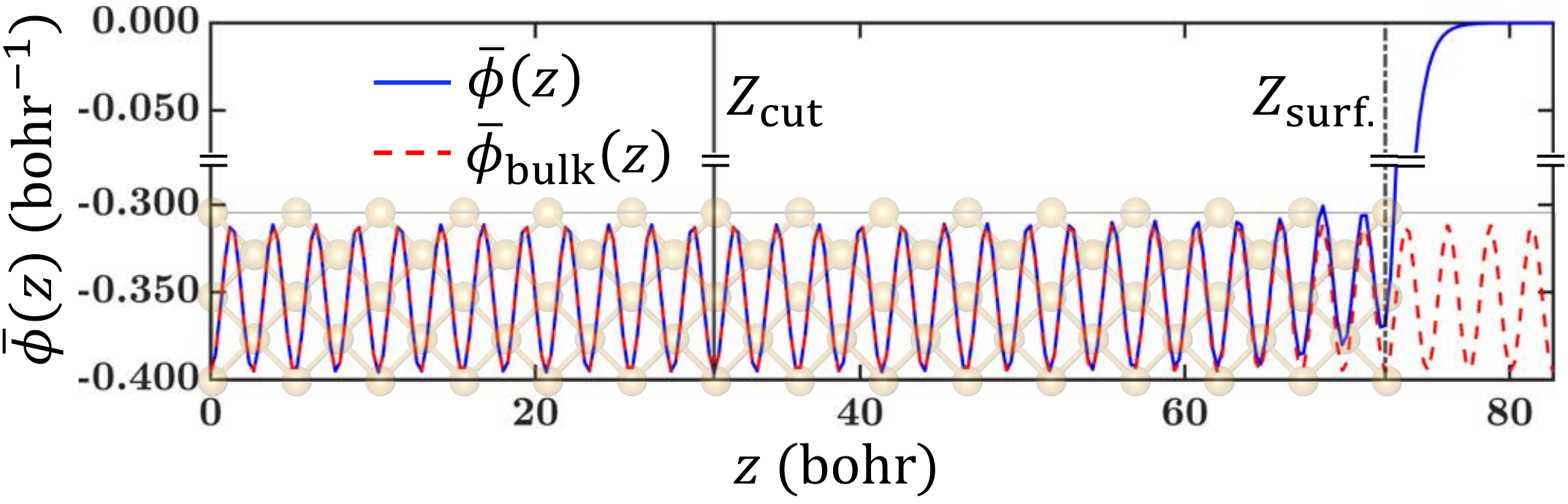}
        \caption{$\bar{\phi}(z)$ for Si}
        \label{fig:Phi_Si}
    \end{subfigure}

    \vspace{1em}

    \begin{subfigure}[b]{0.48\linewidth}
        \centering
        \includegraphics[width=\textwidth]{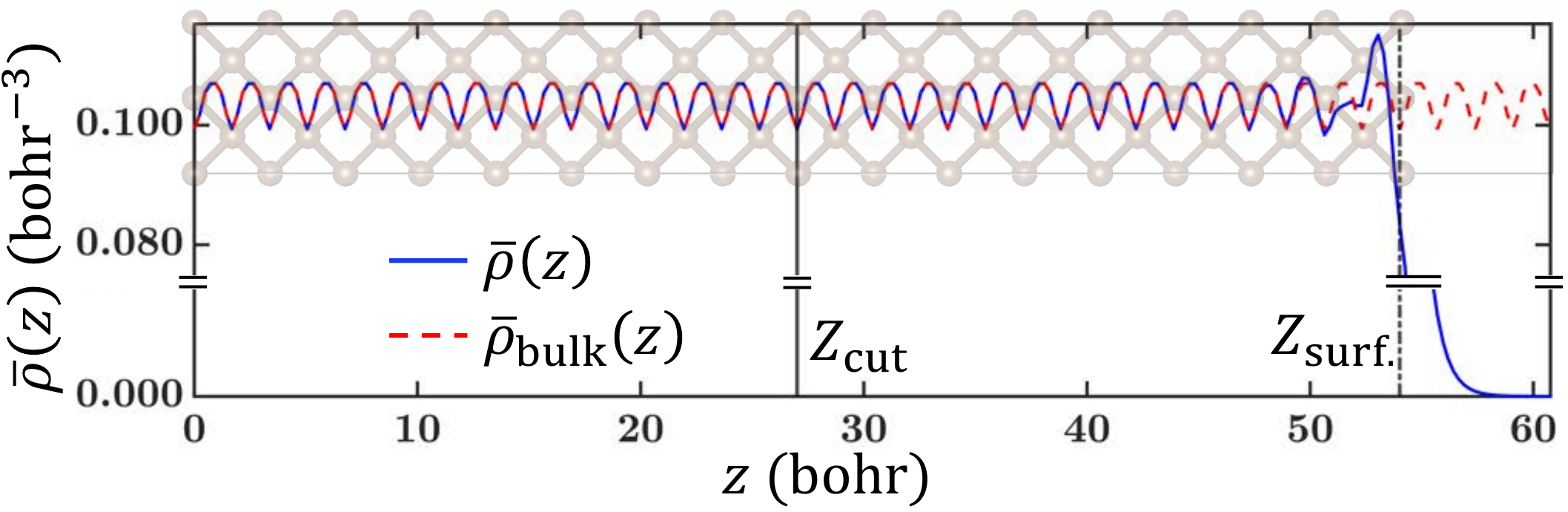}
        \caption{$\bar{\rho}(z)$ for C}
        \label{fig:Rho_C}
    \end{subfigure}
    \hfill
    \begin{subfigure}[b]{0.48\linewidth}
        \centering
        \includegraphics[width=\textwidth]{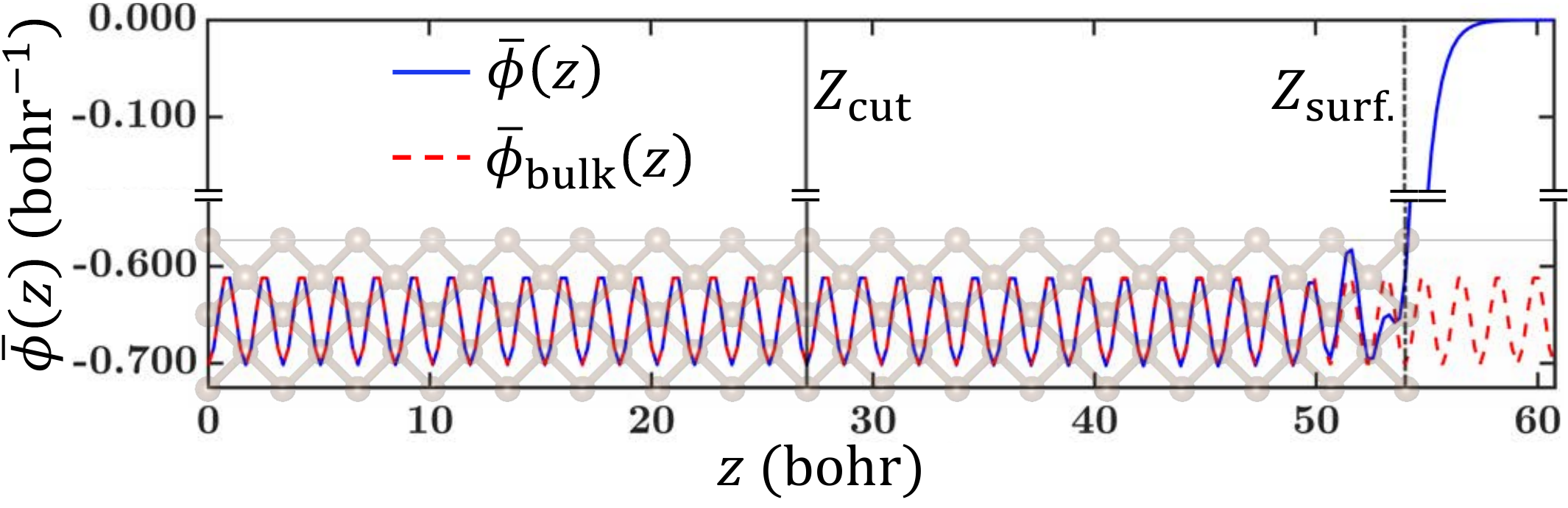}
        \caption{$\bar{\phi}(z)$ for C}
        \label{fig:Phi_C}
    \end{subfigure}

    \caption{Convergence of (a) the electron density and (b) the electrostatic potential to their bulk counterparts as a function of unit cell depth from the surface, measured as $\Delta\rho = \|\rho_{\mathrm{cell}} - \rho_{\mathrm{bulk}}\| / \|\rho_{\mathrm{bulk}}\|$ and $\Delta\phi = \|\phi_{\mathrm{cell}} - \phi_{\mathrm{bulk}}\| / \|\phi_{\mathrm{bulk}}\|$, respectively.  (c)--(h) Planar-averaged electron density $\bar{\rho}(z)$ (left column) and electrostatic potential $\bar{\phi}(z)$ (right column) computed using the bulk boundary condition formalism (solid blue) and the 3D bulk reference (dashed red), for (c,d)~Al, (e,f)~Si, and (g,h)~C. The domain $\Omega$ consists of four unit cells.}
    \label{fig:Rho_and_Phi_combined}
\end{figure}

We next examine the convergence of the electron density and electrostatic potential toward their bulk counterparts with increasing depth from the surface. In particular, we consider four unit cells in $\Omega$ for each of the Al, Si, and C systems. For each unit cell below the surface, we define the relative errors $\Delta\rho = \|\rho_{\mathrm{cell}} - \rho_{\mathrm{bulk}}\| / \|\rho_{\mathrm{bulk}}\|$ and $\Delta\phi = \|\phi_{\mathrm{cell}} - \phi_{\mathrm{bulk}}\| / \|\phi_{\mathrm{bulk}}\|$ with respect to the 3D bulk references $\rho_{\mathrm{bulk}}$ and $\phi_{\mathrm{bulk}}$; $\Delta\rho$ is computed only within $\Omega$, since $\rho = \rho_{\mathrm{bulk}}$ for $z \leq Z_{\mathrm{cut}}$ by construction. As shown in Figs.~\ref{fig:Rho_convergence} and \ref{fig:Phi_convergence}, both $\Delta\rho$ and $\Delta\phi$ decrease systematically and monotonically with depth, the behavior of $\Delta\phi$ also validating the numerical treatment of the undetermined constant in $\phi_{\mathrm{bulk}}$. The convergence is slower for Al than for Si and C, as expected from the slower decay of the density matrix in Al, reflecting the longer-ranged electronic interactions in metallic systems. The planar-averaged profiles of the electron density (Figs.~\ref{fig:Rho_Al}, \ref{fig:Rho_Si}, and \ref{fig:Rho_C}) and electrostatic potential (Figs.~\ref{fig:Phi_Al}, \ref{fig:Phi_Si}, and \ref{fig:Phi_C}) further confirm agreement with the bulk reference throughout the bulk-like region, with deviations confined to the vicinity of the surface, as expected.

\begin{figure}[h!]
    \centering
    \begin{subfigure}[b]{0.38\linewidth}
        \centering
        \includegraphics[width=\textwidth]{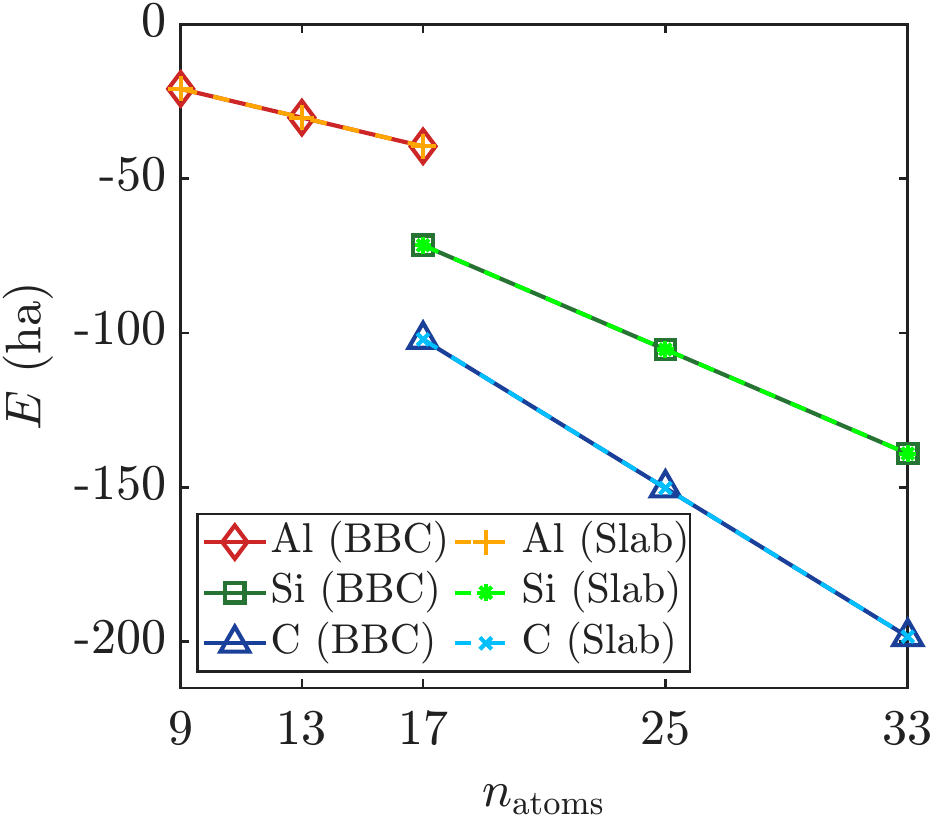}
        \caption{Energy}
        \label{fig:Linear_fit_relaxed}
    \end{subfigure}
    \hspace{1em}
    \begin{subfigure}[b]{0.38\linewidth}
        \centering
        \includegraphics[width=\textwidth]{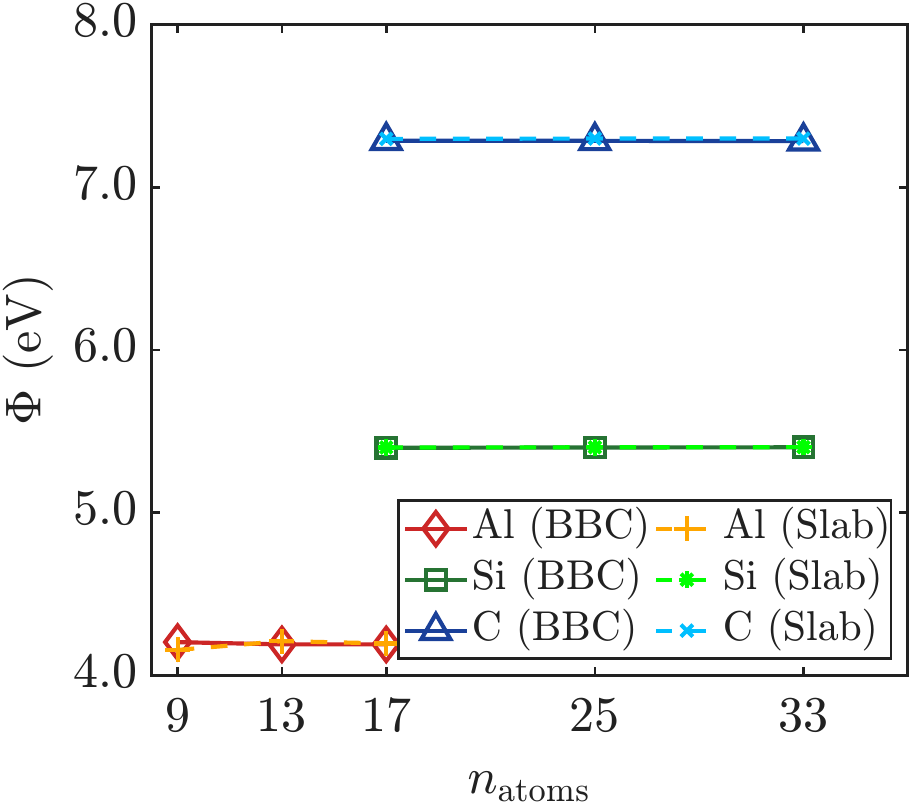}
        \caption{Work Function}
        \label{fig:Work_fn_relaxed}
    \end{subfigure}
    \caption{(a) Linear fit of the total energy $E$ as a function of the number of atoms $n_{\mathrm{atoms}}$ for the (100) surfaces, obtained using the bulk boundary condition (BBC) and slab formalisms. (b) Variation of the work function $\Phi$ of the (100) surfaces with $n_{\mathrm{atoms}}$. To enable direct comparison with the BBC results, both $E$ and $n_{\mathrm{atoms}}$ for the slab are halved, as permitted by its inversion symmetry.}
    \label{fig:E_and_work_fn}
\end{figure}

We now calculate the surface energies using the bulk boundary condition method, and compare against those obtained with the slab formalism. The surface energy is defined as the excess energy arising from the creation of a surface. Accordingly, the total energy is fit to the form:
\begin{equation}
    E = \begin{cases} n_{\mathrm{atoms}}\, \varepsilon_{\mathrm{bulk}} + n_{\mathrm{surf}}\, \varepsilon_{\mathrm{surf}} & \text{(bulk boundary condition)} \\[4pt] n_{\mathrm{atoms}}\, \varepsilon_{\mathrm{bulk}} + 2\, n_{\mathrm{surf}}\, \varepsilon_{\mathrm{surf}} & \text{(slab)} \end{cases}
\end{equation}
where $n_{\mathrm{atoms}}$ is the number of atoms in $\Omega$, $\varepsilon_{\mathrm{bulk}}$ is the bulk energy per atom, $n_{\mathrm{surf}} = 2$ is the number of surface atoms, the factor of 2 in the slab case accounts for its two symmetric surfaces, and $\varepsilon_{\mathrm{surf}}$ is the surface energy per surface atom. In particular, $\varepsilon_{\mathrm{bulk}}$ is obtained from the slope of a linear fit of $E$ against $n_{\mathrm{atoms}}$ at varying domain sizes, with $\varepsilon_{\mathrm{surf}}$ following from the intercept---an approach that yields well-converged surface energies independent of the bulk energy reference~\cite{Fiorentini_1996, patra2017}. In Fig.~\ref{fig:Linear_fit_relaxed}, we plot $E$ against $n_{\mathrm{atoms}}$ for the bulk boundary condition and slab formalisms. The data points from the two methods fall on top of each other for all three materials, with $R^2 = 1$ for all linear fits, and the bulk energy per atom extracted from the fitted slopes agrees with the 3D bulk reference to within $\mathcal{O}(10^{-5})$ ha/atom, confirming the accuracy of the procedure. The corresponding surface energies, reported in Table~\ref{tab:bulk_vs_slab}, agree to within $0.02$~eV/surface atom for all three materials, i.e., to within the accuracy of the calculations themselves.

We next calculate the work function $\Phi$, the minimum energy required to remove an electron from the solid through the given surface to the vacuum: $\Phi = \phi_{\mathrm{vac}} - \mu$, where $\phi_{\mathrm{vac}}$ is the electrostatic potential in the vacuum region and $\mu$ is the Fermi level. In the bulk boundary condition formalism, $\phi_{\mathrm{vac}} = 0$ by the Dirichlet boundary condition imposed at the vacuum boundary, while in the slab approximation it follows from the inversion symmetry of the geometry. The results are shown in Fig.~\ref{fig:Work_fn_relaxed} as a function of domain size. Both methods converge rapidly for Si and C, reaching convergence at the smallest domain considered. For Al, the formalisms requires modest additional depth, consistent with its metallic character and the associated Friedel oscillations~\cite{BeFriedel1997, Schultz2021}. At the largest domain size, the two methods agree to within $0.02$~eV, as shown in Table~\ref{tab:bulk_vs_slab}. The consistent agreement in surface energies and work functions across this range of systems verifies the accuracy of the proposed bulk boundary condition formalism.

\begin{table}[h!]
    \centering
    \caption{Comparison of the (100) surface energies $\varepsilon_{\mathrm{surf}}$ (eV/surface atom) and work functions $\Phi$ (eV) of Al, Si, and C, obtained using the bulk boundary condition (BBC) and slab formalisms.}
    \label{tab:bulk_vs_slab}
    \begin{tabular}{lccc}
        \toprule
        & Method & $\varepsilon_{\mathrm{surf}}$ & $\Phi$ \\
        \midrule
        \multirow{2}{*}{Al (100)} & BBC & 0.484 & 4.191 \\
         & Slab & 0.505 & 4.197 \\
        \midrule
        \multirow{2}{*}{Si (100)} & BBC & 2.012 & 5.402 \\
         & Slab & 2.022 & 5.403 \\
        \midrule
        \multirow{2}{*}{C (100)} & BBC & 3.370 & 7.283 \\
         & Slab & 3.385 & 7.299 \\
        \bottomrule
    \end{tabular}
\end{table}


Finally, we study CO adsorption on Cu(100) at 0.5~ML coverage at the top (T), bridge (B), and hollow (H) sites. As before, we employ the PBE exchange-correlation functional and ONCV pseudopotentials from the SPMS set, with the computational domain spanning a single unit cell in the lateral directions. The equilibrium lattice constant of bulk Cu is $a_\text{lat} = 6.84$~bohr, and the gas phase equilibrium bond length of CO is $d_\text{C-O} = 2.141$~bohr. In the bulk boundary condition formalism, $\Omega$ contains 7 layers and $Z_\text{cut} = 2\,a_\text{lat}$, with atomic positions taken from the topmost layers of the relaxed clean slab. In the slab formalism, the slab comprises 13 layers and is relaxed with the top and bottom two layers free and inversion symmetry enforced. In both formalisms, a vacuum of $13.68$~bohr is employed. In the bulk boundary condition formalism, a single CO molecule is placed on the exposed surface, while in the slab formalism, two CO molecules are placed symmetrically on the opposite slab faces to enforce inversion symmetry. In both cases, the CO bond length and site-dependent adsorbate--substrate distances are fixed at the values reported by Favot et al.~\cite{COadsFavot2001}: $d_\text{C-O} = 2.173$~bohr and $d_\text{Cu-C} = 3.553$, $2.948$, and $2.287$~bohr for the T, B, and H sites, respectively, where $d_\text{Cu-C}$ denotes the vertical height of C above the surface Cu plane. The adsorption energy is evaluated at the corresponding electronic ground state. In all calculations, we use Fermi--Dirac smearing of 0.1~eV and a grid spacing of $h = 0.27$~bohr, with the surface calculations employing a $5 \times 5$ k-point grid and the corresponding bulk calculations a $5 \times 5 \times 11$ grid. These and other numerical parameters  are chosen such that the adsorption energies are converged to within 0.1~eV/molecule. The adsorption energy per CO molecule is defined as: 
\begin{equation}
    E_\text{ads} = \begin{cases} E_\text{surf+CO} - E_\text{surf} - E_\text{CO} & \text{(bulk boundary condition)} \\[8pt] \dfrac{1}{2}\left(E_\text{surf+2CO} - E_\text{surf} - 2E_\text{CO}\right) & \text{(slab)} \end{cases}
\end{equation}
where $E_\text{surf}$ denotes the total energy of the clean surface system, $E_\text{surf+CO}$ that of the surface with a single adsorbed CO molecule in the bulk boundary condition formalism, $E_\text{surf+2CO}$ that of the slab with CO molecules adsorbed symmetrically on both faces, and $E_\text{CO}$ the total energy of the relaxed gas-phase molecule. The results are reported in Table~\ref{tab:co_adsorption}. The bulk boundary condition adsorption energies agree with the slab values to within $0.05$~eV for the T site and $\sim 0.09$~eV for the B and H sites, i.e., to within the accuracy of the calculations themselves. In addition, the slab adsorption energies of $-0.79$, $-0.70$, and $-0.63$~eV for the T, B, and H sites are in good agreement with the relaxed PBE values of $-0.77$, $-0.72$, and $-0.68$~eV reported by Favot et al.~\cite{COadsFavot2001}. Notably, both formalisms reproduce the site ordering T $>$ B $>$ H.

\begin{table}[h!]
\centering
\caption{CO adsorption energy $E_\text{ads}$ (eV/molecule) on Cu(100) at 0.5~ML coverage for the top (T), bridge (B), and hollow (H) sites, computed using the bulk boundary condition (BBC) and slab formalisms.}
\begin{tabular}{lccc}
\hline
\textbf{Method} & \textbf{T} & \textbf{B} & \textbf{H} \\
\hline
BBC          & $-0.74$ & $-0.61$ & $-0.55$ \\
Slab & $-0.79$ & $-0.70$ & $-0.63$ \\
\hline
\end{tabular}
\label{tab:co_adsorption}
\end{table}

We now discuss the computational cost of the bulk boundary condition formalism relative to the slab formalism. In the examples presented here, the number of layers required in $\Omega$ for the bulk boundary condition formalism is nearly  half that required in the slab formalism. The relative cost is then primarily determined by $Z_{\mathrm{cut}}$, which decreases with increasing band gap in insulating systems and increasing smearing in metallic systems. In practical simulations, $Z_{\mathrm{cut}}$ is expected to correspond to around 3--6 layers, in which case the eigenproblems in the two formalisms are of the same size, making the cost of computing the electronic ground state comparable. The advantages of the bulk boundary condition formalism become apparent in geometry optimization, where the number of atoms to be relaxed is half that in the slab formalism, and the more involved setup of mirroring adsorbates on both slab faces is avoided altogether. These advantages become even more pronounced for systems with longer-ranged electrostatic interactions, e.g., those with large dipole moments, where the number of slab layers must be significantly increased to suppress spurious electrostatic coupling across the slab, whereas in the bulk boundary condition formalism only the electrostatic domain needs to be extended, at negligible additional cost.


In summary, the proposed bulk boundary condition formalism provides an accurate and efficient avenue for surface calculations in Kohn--Sham DFT. There are several natural extensions of the present work. Implementation in the large-scale open-source electronic structure code SPARC~\cite{xu2021sparc, zhang2024sparc} would make the formalism widely accessible. Generalization of the formalism to nonlocal exchange-correlation functionals, including real-space hybrid~\cite{XinHybrid2024} and random-phase approximation~\cite{zhang2025rpa}, would allow for higher fidelity calculations. In addition, employing bulk boundary conditions in the plane parallel to the surface would enable access to the low-coverage limit, while applying them in all three directions would allow the study of bulk defects such as vacancies~\cite{bhowmikDefect2025,ofdft_bulkBC}. The formalism also extends naturally to interfaces between two dissimilar materials~\cite{vanDeWalle1987, Dardzinski_2022}, with bulk boundary conditions applied on both sides of the interface. On the electrochemistry front, incorporating a fixed electrode potential in the bulk would enable constant-potential simulations~\cite{SUNDARARAMAN2017278, Kastlunger2018}, and the inclusion of explicit or implicit solvation models~\cite{Hormann2019,Gauthier2019} would further widen the applicability of the formalism to electrochemical systems. Finally, accommodating charged systems~\cite{Chan2015,AKHADE201763} would broaden its scope to a even wider range of problems in surface science.


\section*{Acknowledgements}
The authors gratefully acknowledge the support of the U.S. Department of Energy, Office of Science, under Grant No. DE-SC0023445. This research was also supported by the supercomputing infrastructure provided by Partnership for an Advanced Computing Environment (PACE) through its Phoenix cluster at Georgia Institute of Technology, Atlanta, Georgia.


\bibliography{bibliography}

\end{document}